\documentclass[10pt,twocolumn]{ICCAS}
 
\usepackage{tabu}
\usepackage{diagbox}
\usepackage{cite}
\usepackage{array}
\usepackage{comment}
\usepackage{anyfontsize}
\usepackage{mathtools}
\usepackage{amsmath, dsfont, bm}

\begin{document}

\title{Correlation Analysis Between MF R-Mode Temporal ASF and Meteorological Factors}

\author{Jongmin Park${}^{1}$, Junwoo Song${}^{1}$, Taewon Kang${}^{1}$, Jaewon Yu${}^{1}$, and Pyo-Woong Son${}^{2*}$ }

\affils{ ${}^{1}$School of Integrated Technology, Yonsei University, \\
Incheon, 21983, Korea (jm97, junwoo3529, taewon.kang, jaewon.yu@yonsei.ac.kr) \\
${}^{2}$Department of Electronics Engineering, Chungbuk National University, \\
Cheongju, 28644, Korea (pwson@cbnu.ac.kr) \\
{\small${}^{*}$ Corresponding author}}

%\thanks{ \noindent
%   This paper is supported by my funding agencies.
%  }

\abstract{
As the vulnerabilities of global navigation satellite systems (GNSS) have become more widely recognized, the need for complementary navigation systems has grown. 
Medium frequency ranging mode (MF R-Mode) has gained attention as an effective backup system during GNSS outages, owing to its strong signal strength and cost-effective scalability.
However, to achieve accurate positioning, MF R-Mode requires correction for the additional secondary factor (ASF), a propagation delay affected by terrain. 
The temporal variation of ASF, known as temporal ASF, is typically corrected using reference stations; however, the effectiveness of this method decreases with distance from the reference station.
In this study, we analyzed the correlation between temporal ASF and meteorological factors to evaluate the feasibility of predicting temporal ASF based on meteorological factors.
Among these factors, temperature and humidity showed significant correlations with temporal ASF, suggesting their potential utility in ASF correction.
}


\maketitle

%-----------------------------------------------------------------------

\section{Introduction}

As the vulnerabilities of global navigation satellite systems (GNSS) \cite{Chiou08:Performance, Sun21:Markov, Chen10:Real, DeLorenzo10:WAAS/L5}, such as jamming and spoofing, have become increasingly recognized, the need for complementary navigation systems has garnered growing attention \cite{Park18:Dual, Park21:Single, Lee19:Development, KIM19:Mitigation, Lee22:Urban, Lee22:Optimal, Park24:CSAC, Kim23:Low, Lee23:Seamless, Kim22:Machine}. 
Due to their strong signal strength, terrestrial radio navigation systems have emerged as promising backup solutions during GNSS outages.
Representative terrestrial systems include medium frequency ranging mode (MF R-Mode), very high frequency data exchange system ranging mode (VDES R-Mode), the enhanced long-range navigation system (eLoran), and DME/DME, which utilizes distance measurement equipment (DME) \cite{Jeong21:Development, Yu22:Simulation, Grundhofer22:Phase, Hehenkamp24:Prediction, Wirsing23:Direct, Son18:Novel, Wirsing25:VDES, Son18:Preliminary, Rhee21:Enhanced, Son24:eLoran, Kim17:SFOL, Lee22:SFOL, Park25:Toward}.
In addition, RSS-based opportunistic navigation and target localization methods, as well as shadow matching and ray tracing techniques, have also been explored \cite{Moon24:HELPS, Lee22:Performance, Kim25:Set, Lee25:Reducing, Kim24:Performance, Jeong24:Quantum, Lee24:Efficient, Kim23:Single, Lee23:Performance_Comparison, Kim23:Machine, Lee23:Performance_Evaluation, Lee22:Evaluation}.

Among these, MF R-Mode operates in the medium frequency (MF) band ranging from 283.5 to 325 kHz and estimates position by measuring the phase of continuous wave (CW) signals transmitted from ground stations \cite{Grundhofer22:Phase, Hehenkamp24:Prediction}. One of its advantages is that existing differential GPS (DGPS) stations can be retrofitted with minimal modifications to function as MF R-Mode transmitters, enabling cost-effective scalability \cite{Johnson14:Feasibility3, Johnson20:R-Mode}.

However, because MF R-Mode relies on ground wave propagation, it is subject to propagation delay influenced by terrain, which must be corrected to achieve accurate positioning. 
The delay consists of three components: the primary factor (PF) caused by atmospheric effects, the secondary factor (SF) caused by the sea surface, and the additional secondary factor (ASF) caused by the terrain surface \cite{Son19:Universal, Kim22:First, Liu23:ELoran}. 
While PF and SF can be corrected using existing models (e.g., Brunavs' equation), ASF is difficult to model due to its high spatial and temporal variability \cite{Son19:Universal}.

ASF is generally divided into a spatial ASF, which remains constant over time and a temporal ASF, which varies over time.
Temporal ASF is typically corrected using reference stations, but its correction accuracy decreases with distance from the reference station.

To address this limitation, recent studies in the eLoran system have explored the use of deep learning techniques to predict temporal ASF based on meteorological factors \cite{Pu21:Accuracy, Liu23:ELoran, Kang25:Enhancing}. 
Inspired by this approach, the present study investigates the correlation between temporal ASF of the MF R-Mode  system and meteorological factors using the Pearson correlation coefficient (PCC) to examine the feasibility of weather-based temporal ASF prediction.

The remainder of this paper is structured as follows.
Section 2 introduces the datasets used for the analysis, including both temporal ASF and meteorological measurements.
Section 3 presents the results of the correlation analysis, followed by a discussion of the findings and potential limitations.
Section 4 concludes the paper and suggests directions for future research.

\section{Data and methodology}

\begin{figure} 
    \centering 
    \includegraphics[width=1.0\linewidth]{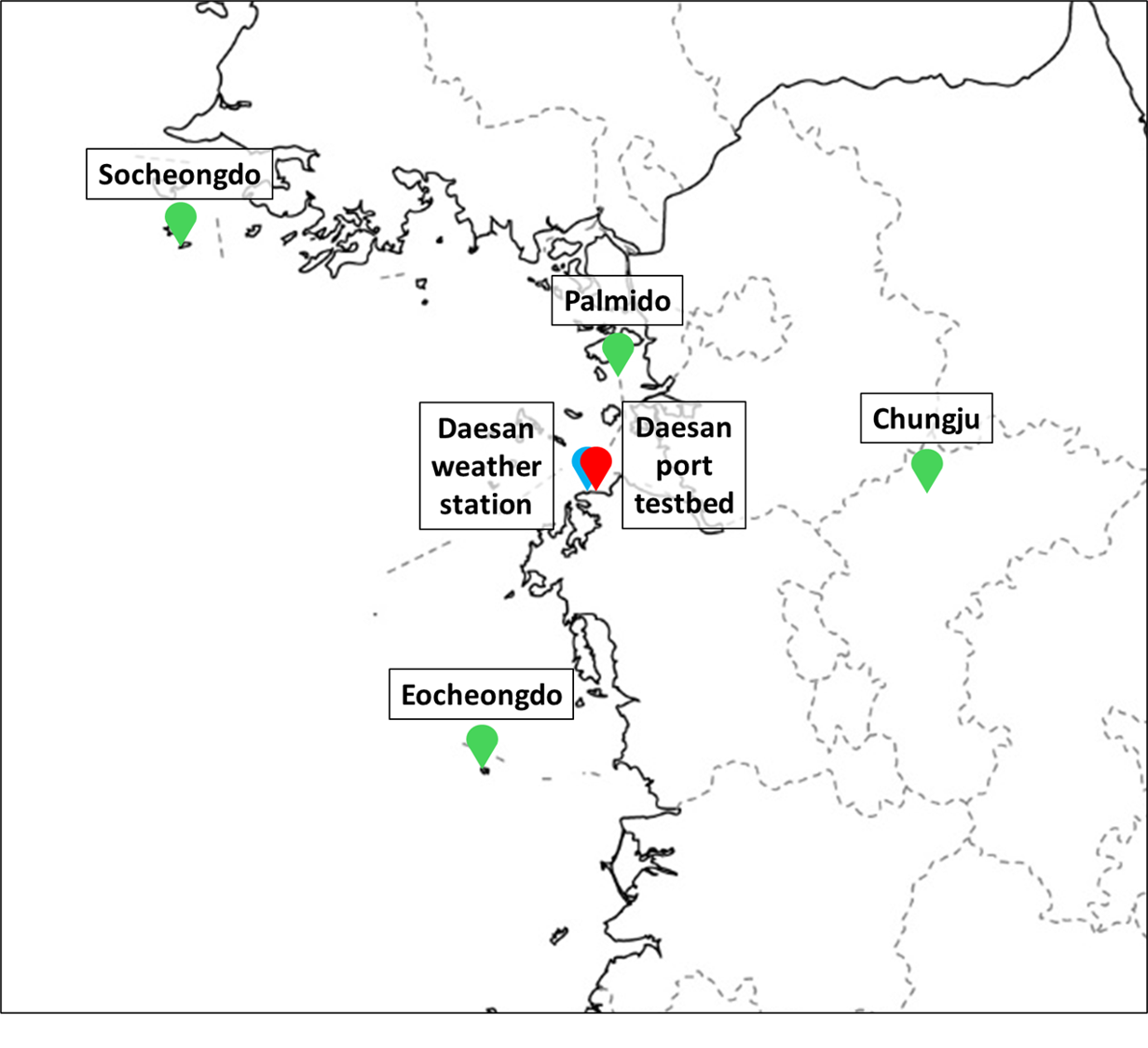}
    \caption{The locations of the testbed, weather station, and transmitters.} 
    \label{fig:locations}
\end{figure}

The data used for the correlation analysis include the temporal ASF measured at the Daesan port testbed (N 37.01279, E 126.41891) and the meteorological measurement collected at the Daesan weather station (N 37.01061, E 126.38808), which is located approximately 2.7 km from the testbed. 
The MF R-Mode transmitting stations are located in Chungju (CJ), Eocheongdo (EC), Socheongdo (SC), and Palmido (PM), South Korea. 
The locations of the testbed, weather station, and transmitters are shown in Fig.~\ref{fig:locations}.

\subsection{Temporal ASF data}
Temporal ASF was calculated using MF R-Mode phase measurements recorded at 1-second intervals collected on ten days: May 30–31 and June 2, 5, 8, 9, 21, 26–28, 2023. 
Only daytime data (06:00–18:00) were used to minimize the influence of skywave propagation, which is typically more pronounced at night and may introduce multipath interference \cite{Jeong22:Preliminary}.

The procedure for calculating the temporal ASF is described in the following equations. 
First, the MF R-Mode phase measurements, sampled at 1-second intervals, were averaged over non-overlapping 5-minute windows to reduce short-term noise and emphasize slow temporal variations.
This 5-minute average phase value for the \textit{i}th interval is given by:
\begin{equation}
\label{eqn:tempASF1}
\bar{\sigma_i} = \frac{1}{300} \sum_{t=300(i-1)+1}^{300i} \sigma(t)
\end{equation}
where $\sigma(t)$ denotes the phase measurement at time $t$.
Next, the overall mean of all phase measurements was calculated as:
\begin{equation}
\label{eqn:tempASF2}
\bar{\sigma}_{\text{total}} = \frac{1}{N} \sum_{t=1}^{N} \sigma(t)
\end{equation}
where $N$ is the total number of phase measurements.
The temporal variation of the phase was calculated as:
\begin{equation}
\label{eqn:tempASF3}
\Delta \sigma_i = \bar{\sigma_i} - \bar{\sigma}_{\text{total}}
\end{equation}

This step isolates the slow temporal changes in phase over time by removing the static component.
Finally, the phase variation was converted into a time delay in nanoseconds to derive the temporal ASF.
The conversion from phase to temporal ASF is expressed by:
\begin{equation}
\label{eqn:tempASF4}
\text{ASF}_{i} = \frac{\Delta \sigma_i}{2\pi f} \cdot 10^9
\end{equation}
where $f$ is the MF R-Mode carrier frequency (283.5–325 kHz).

\subsection{Meteorological data}
The Daesan weather station records temperature, wind velocity, humidity, and atmospheric pressure at 1-minute intervals. 
The measurements were obtained from the Korea Meteorological Administration (KMA). 
The meteorological measurements were also averaged over 5-minute intervals to match the temporal resolution of the ASF data.

\subsection{Pearson correlation coefficient and \textit{p}-value}

The PCC ranges from $-1$ to $1$ and quantifies the strength and direction of the linear relationship between two continuous variables. 
A value close to $1$ indicates a strong positive correlation, while a value close to $-1$ indicates a strong negative correlation. 
The \textit{p}-value assesses the statistical significance of the PCC, with values below $0.05$ generally considered significant.

\section{Analysis results and discussion}

\begin{figure} [bp!]
    \centering 
    \includegraphics[width=1.0\linewidth]{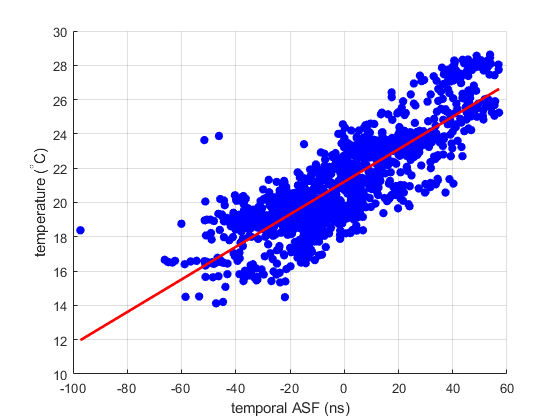}
    \caption{Correlation between temperature and temporal ASF.} 
    \label{fig:result_temp}
\end{figure}

\begin{figure} [bp!]
    \centering 
    \includegraphics[width=1.0\linewidth]{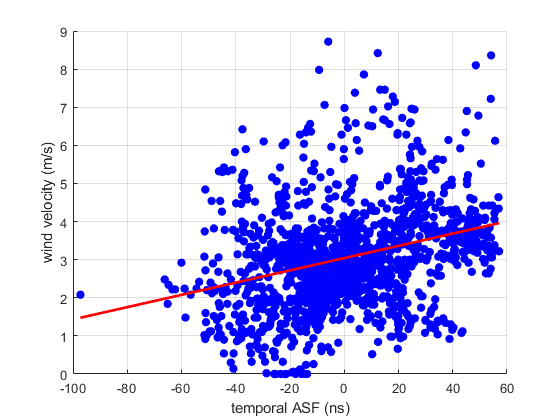}
    \caption{Correlation between wind velocity and temporal ASF.} 
    \label{fig:result_wind}
\end{figure}

\begin{figure} [bp!]
    \centering 
    \includegraphics[width=1.0\linewidth]{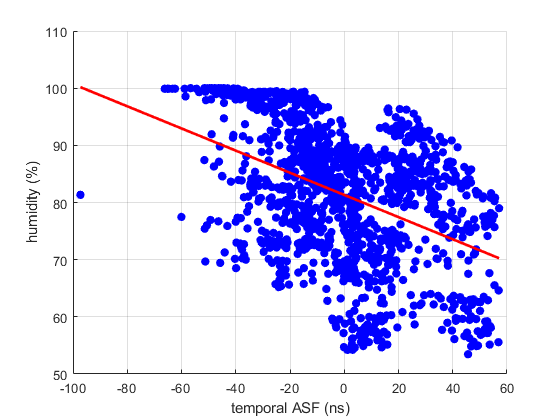}
    \caption{Correlation between humidity and temporal ASF.} 
    \label{fig:result_humid}
\end{figure}

\begin{figure} [bp!]
    \centering 
    \includegraphics[width=1.0\linewidth]{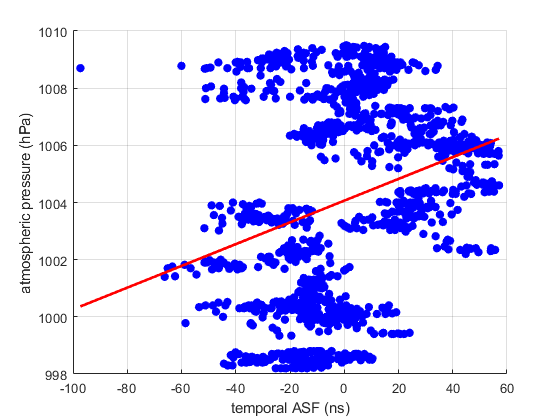}
    \caption{Correlation between atmospheric pressure and temporal ASF.} 
    \label{fig:result_press}
\end{figure}

All correlation results had \textit{p}-values below 0.05, confirming the statistical significance of the correlation coefficient.
The PCCs calculated for each transmitter are summarized in Table.~\ref{tab:PCC}, and the correlations between temporal ASF from the Chungju transmitter and each meteorological factor (temperature, wind velocity, humidity, and atmospheric pressure) are presented in Figs.~\ref{fig:result_temp}, \ref{fig:result_wind}, \ref{fig:result_humid}, and \ref{fig:result_press}.

\begin{table}[bp!]
\centering
\caption{Pearson correlation coefficients between temporal ASF and each meteorological factor}
\label{tab:PCC}
\scriptsize
\begin{tabular}{|l|c|c|c|c|} % l=left, c=center, r=right alignment
\hline
Meteorological Factors &   CJ   &  EC  &  SC  &  PM    \\\hline
Temperature ($^\circ\mathrm{C}$)& 0.85 & 0.92 & 0.87 & 0.72 \\\hline
Wind Velocity (m/s)& 0.31 & 0.29 & 0.30 & 0.22\\\hline
Humidity (\%) & -0.43 & -0.59 & -0.40 & -0.46 \\\hline
Atmospheric Pressure (hPa) & 0.28 & 0.50 & 0.21 & 0.26\\\hline
\end{tabular}
\end{table}

The correlation between temporal ASF and meteorological factors revealed clear patterns.
Temperature exhibited a strong positive correlation with temporal ASF, indicating that higher temperatures were consistently associated with increased ASF values.
Humidity exhibited a weak to moderate negative correlation, suggesting that ASF values tended to decrease slightly as humidity increased.
In contrast, both wind velocity and atmospheric pressure demonstrated negligible correlations, implying that their influence on temporal ASF was minimal.

However, this analysis is limited to data from May and June and may not sufficiently reflect seasonal meteorological variations.
Additionally, since the testbed and weather station are located on the coast and tend to have high relative humidity, the characteristics of ASF changes in low-humidity environments may not have been fully captured.

\section{Conclusion}

This study analyzed the relationship between temporal ASF in the MF R-Mode system and meteorological factors and found a strong correlation particularly with temperature.
Humidity showed a moderate correlation, while wind velocity and atmospheric pressure showed relatively weak correlations.

Based on these results, it appears feasible to construct a model to predict MF R-Mode temporal ASF using weather information, especially temperature. 
Such a model could contribute to improving the accuracy and reliability of MF R-Mode positioning, particularly in GNSS-denied environments.

Future work will focus on generalizing the model by incorporating data from diverse seasons and geographical locations.

\section*{ACKNOWLEDGEMENT}

This work was supported in part by Grant RS-2024-00407003 from the ``Development of Advanced Technology for Terrestrial Radionavigation System'' project, funded by the Ministry of Oceans and Fisheries, Republic of Korea; 
in part by the National Research Foundation of Korea (NRF), funded by the Korean government (Ministry of Science and ICT, MSIT), under Grant RS-2024-00358298; 
in part by the Korea Aerospace Administration (KASA), under Grant RS-2022-NR067078; 
in part by the Unmanned Vehicles Core Technology Research and Development Program through the NRF and the Unmanned Vehicle Advanced Research Center (UVARC), funded by the MSIT, Republic of Korea, under Grant 2020M3C1C1A01086407; 
in part by the Institute of Information \& Communications Technology Planning \& Evaluation (IITP) through the Information Technology Research Center (ITRC) program, funded by the MSIT, under Grant IITP-2025-RS-2024-00437494.

Generative AI (ChatGPT, OpenAI) was used solely to assist with grammar and language improvements during the manuscript preparation process.  
No content, ideas, data, or citations were generated by AI.  
All technical content, methodology, analysis, and conclusions were written and verified solely by the authors.

\bibliographystyle{IEEEtran}
\bibliography{mybibfile, IUS_publications}

\end{document}